\documentclass[aps,prl,superscriptaddress,twocolumn,showpacs]{revtex4}
\usepackage{graphicx,bm}

\begin{document}

\title{Mapping the Fermi velocity in the quasi-2D
organic conductor $\kappa$-(BEDT-TTF)$_2$I$_3$}

\author{A. E. Kovalev}
\affiliation{Department of Physics, University of Florida,
Gainesville, FL 32611, USA}
\author{S. Hill} \email{hill@phys.ufl.edu} \affiliation{Department of Physics, University of Florida, Gainesville,
FL 32611, USA}
\author{K. Kawano}
\affiliation{Toho University, Funabashi, 274-8510, Japan}
\author{M. Tamura}
\altaffiliation{New address: RIKEN, Wako, Saitama 351-0198,
Japan}\affiliation{Toho University, Funabashi, 274-8510, Japan}
\author{T. Naito}
\affiliation{Hokkaido University, Sapporo, 060-0810, Japan}
\author{H. Kobayashi}
\affiliation{Insitute for Molecular Science, Okazaki, Aichi
444-8585, Japan}

\date{\today}

\begin{abstract}
We demonstrate a new method for determining the Fermi velocity in
quasi-two-dimensional (Q2D) organic conductors. Application of a
magnetic field parallel to the conducting layers results in
periodic open orbit quasiparticle trajectories along the Q2D Fermi
surface. Averaging of this motion over the Fermi surface leads to
a resonance in the interlayer microwave conductivity. The
resonance frequency is simply related to the extremal value of the
Fermi velocity perpendicular to the applied field. Thus, angle
dependent microwave studies enable a complete mapping of the Fermi
velocity. We illustrate the applicability of this method for the
highly-2D organic conductor $\kappa$-(BEDT-TTF)$_2$I$_3$.
\end{abstract}


\maketitle

Microwave spectroscopy has been utilized as a means of studying
the electrodynamic properties of metals for well over half a
century, especially resonant absorption in an external DC magnetic
field. Quasiparticles in such conductors usually move on closed
periodic trajectories in reciprocal ($k$-) space, or cyclotron
orbits in real space (we will always consider the motion in
$k$-space, unless stated otherwise). When any period associated
with this motion matches the period of the external
electromagnetic field, so-called cyclotron resonance (CR) occurs
if the condition $\omega_c\tau>1$ is satisfied, where $\omega_c$
is the cyclotron frequency and $\tau$ is the relaxation time;
$\omega_c$ depends on the magnetic field strength, and on the
cyclotron mass ($m_c$) $-$ a characteristic of the Fermi surface
(FS) of the metal. In conventional metals, large Fermi velocities
($v_F \sim 10^6$~m/s) complicate matters to the extent that CR is
normally only observed in the anomalous skin effect regime. The
theoretical foundations for this and many other electrodynamic
properties of metals have been firmly established
\cite{Abrikosov}.


In layered organic conductors, the FS may be either
quasi-two-dimensional (Q2D), quasi-one-dimensional (Q1D), or a
combination of both. In the Q2D case, the FS is a warped cylinder
with its axis perpendicular to the layers (see Fig.~1) while, in
the Q1D case, the FS consists of a pair of warped sheets at $\pm
k_F$. Because of this reduced dimensionality (and reduced $v_F
\sim 10^5$~m/s) several new effects in the microwave conductivity
have been reported. One of these is the observation of multiple
periodic orbit resonances (POR) in Q1D
systems~\cite{HillPRB97,BlundellPRB97,Ardavan98,KovalevPRB02}. In
this situation, quasiparticles move under the external magnetic
field along open orbits. The motion is periodic because of the
underlying periodicity of the crystal lattice. When the period of
this motion coincides with the period of an appropriately
polarized electromagnetic field, resonant microwave absorption
occurs, {\em i.e.} the AC conductivity attains a maximum. Recent
studies have shown that multiple POR harmonics may be observed,
each one corresponding to a different fourier component of the
warping of the Q1D FS~\cite{KovalevPRB02}. The POR frequencies are
simply related to the magnetic field strength, the crystal lattice
parameters, and the Q1D Fermi velocity ($v_F$). Indeed, this
effect provides the most direct method of measuring $v_F$ in a Q1D
system, without making any assumptions about the underlying
bandstructure~\cite{KovalevPRB02}.

Although a form of closed orbit CR has been reported for several
Q2D organic conductors, its origin is quite different from that in
normal metals~\cite{HillPRB97,McKenziePRB99}. Electromagnetic
fields penetrate easily into the bulk of the sample for current
excitation normal to the layers~\cite{HillPRB00}, due to the very
low conductivity in this direction (the anomalous skin effect is
impossible to achieve in the millimeter spectral range, even for
the highest conducting direction); the typical interlayer skin
depth at 50 GHz is about $50-100~\mu$m, {\em i.e.} comparable with
the sample dimensions. Indeed, typical sample shapes and
conductivity anisotropies result in a situation wherein the
interlayer conductivity ($\sigma_{zz}$) usually dominates the
electrodynamic response~\cite{HillPRB00}. Thus, it has been shown
both theoretically~\cite{HillPRB97,McKenziePRB99} and
experimentally that the Q2D closed orbit resonances are related to
the finite warping of the FS, {\em i.e.} this effect is analogous
to the Q1D POR, albeit that the periodic motion normal to the
layers is related to the underlying cyclotron motion which is
predominantly confined to within the layers.

A new effect appears if one aligns the magnetic field within the
layers of a Q2D conductor, {\em i.e.} perpendicular to the Q2D FS
cylinder. In this case, the quasiparticle motion is {\it
principally} open (except for a small fraction of the total
electrons $-$ see Fig.~1a). This results in periodic motion along
open orbits normal to the layers. The period of this motion
depends on the magnetic field strength, $B$, and the velocity
component ($v_\perp$) perpendicular to the field. Consequently,
the period depends strongly on the in-plane wavevector $k_{xy}$.
However, as we will show, averaging over the FS leads to the
result that the extremal perpendicular velocity ($v_\perp^{ext}$)
dominates the electrodynamic response, giving rise to a
singularity (a form of resonance) in the interlayer conductivity.
The resonance occurs when the period of the electromagnetic field
matches the periodicity of these extremal quasiparticle
trajectories. Thus, one can map out the Fermi velocity by this
method. This effect was originally predicted by Peschanskii and
Pantoja~\cite{Peschansky1}, albeit within the anomalous skin
effect regime. In the following, we derive some expressions for
the real experimental case, where the skin depth is bigger or
comparable to the sample size. Using these calculations, we show
that it is also possible to map out the Fermi velocity in this
limit.

We consider an energy dispersion of the form:


\begin{equation}
\label{parab}
E(\vec{k})=\frac{\hbar^2k_x^2}{2m_x}+\frac{\hbar^2k_y^2}{2m_y}-2t_\perp
\cos(\vec{k}\cdot\vec{R}),
\end{equation}


\noindent{where $m_x$ and $m_y$ are the in-plane diagonal
components of the effective mass tensor, $\vec{R}$ is the real
space vector characterizing the interlayer FS warping, and
$4t_\perp$ is the interlayer bandwidth. For a calculation of the
interlayer AC conductivity [$\sigma_{zz}(\omega)$] we use the
Boltzmann transport equation:}


\begin{eqnarray}
\label{cond1} \sigma_{zz}(\omega)=\frac{e^2}{4\pi^3}\int
d^3\vec{k}\left[-\frac{\partial f_\circ(\vec{k})}{\partial
E(\vec{k})}\right] v_z(\vec{k},0)\nonumber \\
\times\int_{-\infty}^0 v_z(\vec{k},t)e^{-i\omega t}e^{t/\tau}dt,
\end{eqnarray}

\noindent{where $\tau$ is the quasiparticle relaxation time.
First, we assume a T~$=0$ limit, so that the derivative of the
distribution function, $f_\circ(\vec{k})$, may by replaced by
$\delta[E_F-E(\vec{k})]$, where $E_F$ is the Fermi energy. Second,
we neglect the part of the Lorentz force which depends on the
interlayer velocity, $v_z$. Finally, we neglect the effect of the
closed orbits. This assumption is valid provided
$\sqrt{E_F/4t_\perp}~\gg~\tau/T_P$, where $T_P$ is the smallest
period for the open orbits~\cite{Shoenfield,Peschansky2}. From
available Shubnikov-de Haas (SdH) and de Haas-van Alphen (dHvA)
data for the title compound, the ratio $E_F/4t_\perp$ is estimated
to be larger than $10^4$, {\em i.e.} $\kappa$-(BEDT-TTF)$_2$I$_3$
is highly two-dimensional. As we show later, $\tau/T_P$ is of
order unity; thus, the closed orbits may by neglected. As a
result, the interlayer conductivity can be written:}

\begin{equation}
\label{cond2} \sigma_{zz}(\omega)\propto\int_0^{\omega_c^{ext}}
\frac{1}{|v_F|}~\frac{dS_k}{d\omega_c}
~\frac{1-i\omega\tau}{(1-i\omega\tau)^2+(\omega_c\tau)^2}~d\omega_c,
\end{equation}

\noindent{where $dS_k$ is an element on the FS,
$\omega_c=eBav_\perp/\hbar$ is the frequency associated with a
given quasiparticle trajectory on the FS, and $\omega_c^{ext}$ is
the extremal value of $\omega_c$ ($a$ is the interlayer spacing,
and $v_\perp$ is the in-plane velocity perpendicular to the
applied magnetic field).


Looking at Eq.~\ref{cond2}, one can see that $\sigma_{zz}(\omega)$
will be dominated by the extremal perpendicular velocity
$v_\perp^{ext}$ (see Fig.~1b), since $dS_k/d\omega_c$ diverges
({\em i.e.} $d\omega_c/dS_k\rightarrow0$) at these points on the
FS. This leads to a resonance condition whenever
$\omega=\omega_c^{ext}$ ($\equiv eBav_\perp^{ext}/\hbar$),
provided that $\omega\tau>1$. Measurement of $\omega_c^{ext}$, as
a function of the field orientation $\psi$ within the $xy$-plane,
yields a polar plot of $v_\perp^{ext}(\psi)$. The procedure for
mapping the Fermi velocity is then identical to that of
reconstructing the FS of a Q2D conductor from the measured periods
of Yamaji oscillations~\cite{Kartsovnik92}. Analytically, assuming
one can measure $v_\perp^{ext}(\psi)$, it is then possible to
generate the Fermi velocity $v_F(\phi)$ using the following
transformations (see also Fig.~1b):

\begin{eqnarray}
\label{extr}
v_F=\sqrt{(v_\perp^{ext})^2+v_\parallel^2};\>\>\>\>\>\>\>\>\>\>
\phi=\psi+\arctan\left(\frac{v_\perp^{ext}}{v_\parallel}\right);\nonumber
\\v_\parallel=-\frac{dv_\perp^{ext}}{d\psi}.\>\>\>\>\>\>\>\>\>\>\>\>\>\>\>\>\>\>\>\>\>\>\>\>\>\>\>\>\>\>\>\>\>\>\>\>\>\>\>\>
\end{eqnarray}
\smallskip


One may also calculate the conductivity explicitly from
Eq.~\ref{cond2}, for the parabolic energy dispersion given in
Eq.~\ref{parab}:

\smallskip
\begin{widetext}
\begin{equation}
\label{cond3}
\sigma_{zz}(\omega,B,\psi)=\sigma_{zz}(0)\frac{1}{\sqrt{(1-i\omega\tau)^2+(v_{xm}^2
\sin^2\psi + v_{ym}^2 \cos^2\psi) (\frac{eaB}{\hbar}\tau)^2}}.
\end{equation}
\end{widetext}
\smallskip

\noindent{Here, $\vec{B}$=$(B\cos\psi,B\sin\psi,0)$, and $v_{xm}$
$\&$ $v_{ym}$ are the maximal Fermi velocities along $x$ and $y$
respectively. For a magnetic field sweep, the resonance condition
becomes $B_{res}=\hbar\omega/ea\sqrt{(v_{xm}^2 \sin^2\psi +
v_{ym}^2 \cos^2\psi)}$.}

To observe this new type of POR, we chose the highly-2D organic
superconductor $\kappa$-(BEDT-TTF)$_2$I$_3$~\cite{Ishiguro}. Its
FS may be calculated using a 2D tight binding model, resulting in
a network of overlapping Fermi cylinders~\cite{KajitaSSC87}, as
shown in Fig.~1c. The lattice periodicity results in a removal of
the degeneracy at the points where the cylinders cross each other.
Thus, the actual FS consists of a small Q2D pocket (dark shaded
area in Fig.~1c) and a pair of Q1D open sections. However, the
energy barrier separating these bands is relatively small, leading
to magnetic breakdown at fairly low fields. Indeed, SdH and dHvA
measurements ~\cite{Balthes96} are dominated by the large
breakdown orbit (lightly shaded area in Fig.~1c). Meanwhile,
microwave studies with the field oriented perpendicular to the
layers show very clear closed-orbit POR from both Q2D sections of
the FS without any noticeable evidence for the Q1D
part~\cite{Palassis01}.

A small platelet shaped ($\sim0.7\times0.4\times0.12$~mm$^3$)
single crystal of $\kappa$-(BEDT-TTF)$_2$I$_3$ was studied using a
phase sensitive cavity perturbation technique (described
elsewhere~\cite{MolaRSI00}). The sample was placed on the endplate
of a cylindrical TE$_{011}$ cavity, and field rotation in a plane
perpendicular to the axis of the cavity ($\parallel$ highly
conducting $bc$-plane of the sample) was achieved using a 7~T
split-pair superconducting magnet. All measurements were carried
out at T$=4.5$~K, above the superconducting transition temperature
($T_c$=3.5 K), and at a frequency of 53.9 GHz. According to
published resistivity data \cite{KajitaSSC87}, we estimate a
54~GHz skin depth for currents perpendicular (parallel) to the
layers of $53~\mu$m ($4~\mu$m). Thus, we expect $\sigma_{zz}$ to
dominate the losses in the cavity \cite{HillPRB00}, as required in
order to detect the new open orbit resonance.

In Fig.~2 we plot the field dependence of the microwave absorption
and phase shift (solid curves) for several field orientations
($\psi$) within the highly conducting $bc$-plane of the sample. To
understand the shapes of the curves, we treat the sample as a thin
conducting plate of thickness $d$, subjected to a microwave AC
magnetic field polarized within the plane of the plate. In this
situation, the effective polarizability of the sample has the
form:

\begin{equation}
\label{alpha} \alpha(\omega)=\frac{\tanh(kd/2)}{kd/2}-1,
\end{equation}

\noindent{where $k=(-i\omega\sigma\mu_0)^{1/2}$. The absorption is
then proportional to the imaginary part of the $\alpha(\omega)$,
while the phase shift is proportional to the real part. While
Eq.~\ref{alpha} is clearly only approximate, we find excellent
agreement between the predicted and observed behavior; as an
illustration, we have included fits to the $\psi = -54^\circ$ data
in Fig.~2.} The resonances are rather broad (indicated by arrows),
due to the fairly small $\omega\tau$ product ($\sim 2$). While it
is quite difficult to accurately determine the resonance position
from the absorption data, it is relatively easy to do so using the
phase shift.

In the Fig.~3 we plot the experimentally determined
$v_\perp^{ext}(\psi)$ on a polar diagram. These values were
deduced by three different methods: 1) from the maximum in the
phase shift ($\bullet$); 2) from fits to the absorption curves
($\circ$); and 3), from fits to the phase shift curves ($\ast$).
The dashed line is a fit to the points deduced by method 1, and
the solid line is the corresponding Fermi velocity (from
Eq.~\ref{extr}). The value for $v_{xm}$ is $1.3\times 10^5$~m/s
and $v_{ym}$ is $0.62\times10^5$~m/s. This anisotropy is in good
agreement with the known anisotropy of the small Q2D FS for
$\kappa$-(BEDT-TTF)$_2$I$_3$ (dark shaded region in Fig.~1c). If
one assumes a parabolic dispersion (Eq.~\ref{parab}), it is
possible to compare our data with the band parameters determined
for the large Q2D $\beta$-orbit from optical data by Tamura {\em
et al.}~\cite{Tamura91}. In particular, we may estimate the
effective mass along the $c$-direction as $m_c^\beta=\hbar
k_F^c/v_{xm} \sim 2.5 m_e$, which compares to the value of
$2.4m_e$ determined from the optical measurements~\cite{Tamura91}.

Based on the known value, $S_k = 5.5 \times 10^{18}$~m$^{-2}$, for
the area of the small Q2D section of the FS in {\em k}-space, we
may estimate the momentum averaged cyclotron mass from the
relation $m^*$=$\hbar(S_k/S_v)^{1/2}$, where $S_v$ is the area of
the FS in velocity space (Fig.~3). This method gives $m^* \approx
1.7 m_e$, while the experimental value deduced from the SdH and
dHvA effects is $\sim1.9m_e$~\cite{Balthes96}, where $m_e$ is the
free electron mass; the value deduced from earlier closed orbit
POR measurements is $2.2m_e$~\cite{Palassis01}. The mass
anisotropy ($m_x/m_y$) deduced from the present measurements for
the small Q2D FS (dark shaded region in Fig.~1c) is 4. From fits
to the data, we also estimate a relaxation time of $\tau\sim$5~ps,
or $\omega\tau\sim 2$. Thus, our initial assumption to neglect
closed orbits is fully justified.

In principle, one should also expect an open orbit resonance
effect from the Q1D FS sections (see Fig.~1c), as has recently
been reported for several other low-dimensional organic
conductors~\cite{Ardavan98,KovalevPRB02}. However, no clear
evidence for such behavior was found. Indeed, our current and
earlier~\cite{Palassis01} microwave studies clearly show that the
Q2D FS dominates the electrodynamic properties of
$\kappa$-(BEDT-TTF)$_2$I$_3$. This likely suggests that the
warping of Q1D FS is weaker than that of the Q2D section.

Finally, we comment on the effect of a possible misalignment of
the field rotation axis. Following Peschansky and
Kartsovnik~\cite{Peschansky2}, we may assume that the conductivity
will be dominated by the periodic motion along $k_z$, rather than
by any cyclotron motion in the $k_xk_y$-plane, provided
$\sin\theta\ll1/\omega_c^{ext}\tau$ ($\theta$ is the mis-alignment
angle). Clearly, therefore, the accuracy of our sample alignment
is enough to justify our procedure for mapping the Fermi velocity.

In conclusion, we have demonstrated a new type of
magneto-electrodynamic resonance in a layered conductor with a Q2D
FS. The resonance is observed for a magnetic field applied
parallel to the layers, resulting in periodic open quasiparticle
trajectories normal to the layers. The resonance frequency is
simply related to the extremal value of the Fermi velocity
perpendicular to the applied field. Thus, by rotating the field
within the layers, one can map out the angle dependence of the
Fermi velocity. We illustrate the applicability of this method for
the highly-2D organic conductor $\kappa$-(BEDT-TTF)$_2$I$_3$, for
which we obtain FS parameters which are consistent with published
results.


This work was supported by the NSF (DMR0196430, DMR0196461 and
DMR0239481). S. H. would like to thank the Research Corporation
for financial support.


\clearpage

\begin{figure}
\includegraphics[width=0.45\textwidth] {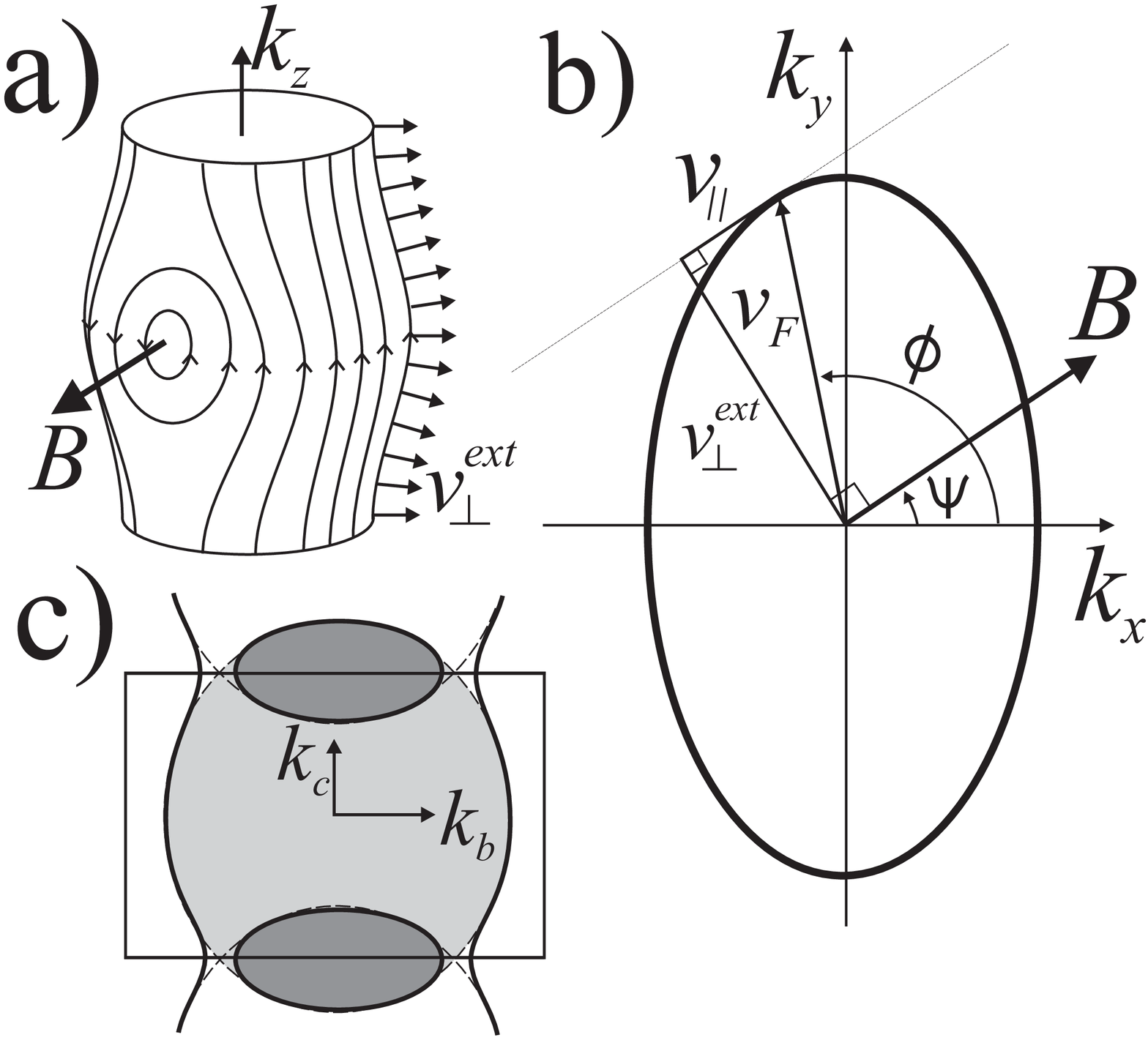}
\caption{\label{cylinder} (a) An illustration of the quasiparticle
trajectories on a warped Q2D FS cylinder for a field oriented
perpendicular to the cylinder axis; the warping has been greatly
exaggerated for clarity. The resulting trajectories lead to a weak
modulation of the quasiparticle velocities parallel to $k_z$ and,
hence, to a resonance in $\sigma_{zz}$. (b) The thick line shows
$v_F(\phi)$ according to Eq.~\ref{parab}; the right angle triangle
illustrates the relationship between $v_\perp^{ext}(\psi)$ and
$v_F(\phi)$ (see text and Eq.~\ref{extr} for explanation). (c) The
Fermi surface of $\kappa$-(BEDT-TTF)$_2$I$_3$ according to
Ref.~\cite{KajitaSSC87}.}
\end{figure}

\begin{figure}
\includegraphics[width=0.5\textwidth]{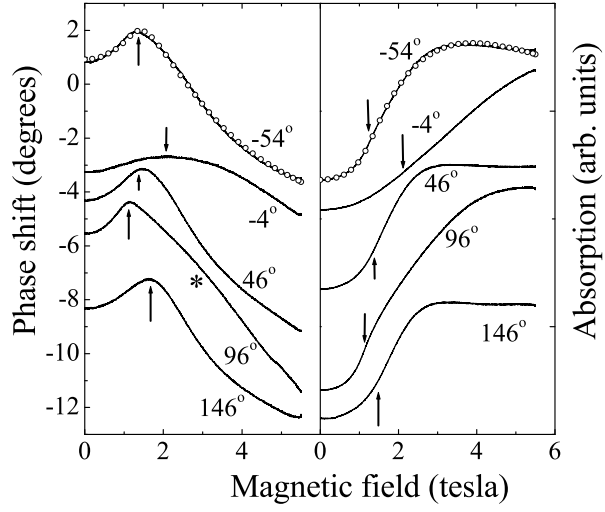}
\caption{\label{traces} Field dependence of the microwave phase
shift (left) and absorption (right) for different field
orientations ($\psi$). The solid curves represent the experimental
data, while fits to the $-54^\circ$ data are shown by open
circles. Arrows mark the resonance positions ($B_{res}$)
determined from fits to Eq.~\ref{cond3}; the asterisk marks a
possible Q1D resonance.}
\end{figure}

\begin{figure}
\includegraphics[width=0.5\textwidth]{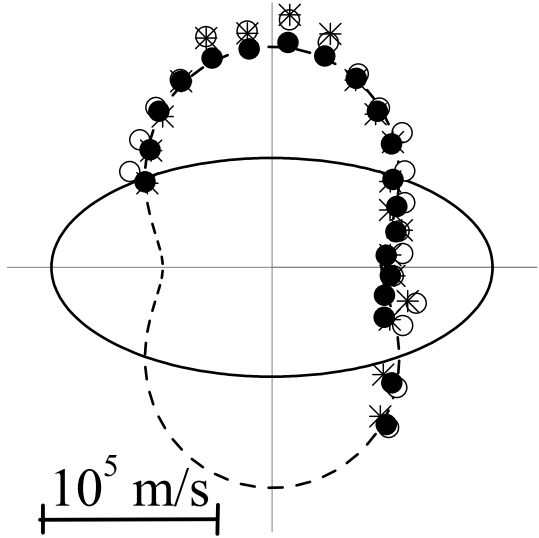}
\caption{\label{velocity}Polar plot of $v_\perp^{ext}(\psi)$
obtained by three different methods: (1) from the maximum phase
shift ($\bullet$); (2) from fits to the absorption traces
($\circ$); and (3), from the fits of the phase shift traces
($\ast$). The dotted line is a fit to the points deduced by
method~1, and the solid line is the resultant Fermi velocity,
$v_F(\phi)$.}
\end{figure}

\end{document}